\documentclass{PoS}
\usepackage{amssymb}
\usepackage{fontenc}
\usepackage{times}
\usepackage{mathptmx}
\usepackage{amsmath}
\usepackage{graphicx}
\usepackage{epsfig}
\usepackage{color}
\let\normalcolor\relax

\newcommand{\ohf}{\ensuremath{\textstyle \frac{1}{2}}}
\newcommand{\thf}{\ensuremath{\textstyle \frac{3}{2}}}

\newcommand{\sugts}{\ensuremath{SU(3)\times\mr{GTS}}}

\newcommand{\ft}{\ensuremath{SU(3)_F\times SU(4)_T}}
\newcommand{\spinks}{\ensuremath{SU(2)_S\times SU(12)_f}}
\newcommand{\su}[2]{\ensuremath{SU(#1)_{#2}}}

\newcommand{\rt}{\ensuremath{\Gamma_4\rtimes SO(4)_T}}
\newcommand{\rrt}{\ensuremath{\Gamma_4\rtimes SU(2)_T}}

\newcommand{\mc}[1]{\ensuremath{\mathcal{#1}}}
\newcommand{\mr}[1]{\ensuremath{\mathrm{#1}}}
\newcommand{\mb}[1]{\ensuremath{\mathbf{#1}}}

\title{Chiral forms and three-flavor operators for staggered baryons}

\ShortTitle{Staggered baryons}

\author{\speaker{Jon A. Bailey}\\
        Washington University, Dept. of Physics - Compton Hall, 1 Brookings Drive - Campus Box 1105, St. Louis, MO  63130, USA\\
        E-mail: \email{jabailey@wustl.edu}}
        
\abstract{In staggered QCD, many staggered baryons correspond to each physical state.  Taste violations lift the continuum degeneracies of the baryons and introduce nonzero off-diagonal elements in the mass matrix.  While presenting no problem of principle, these splittings and mixings complicate analyses of simulation results.  However, in special cases operators with good $SU(3)$ quantum numbers can be used to circumvent the splittings and mixings.  I review what has been learned from staggered chiral perturbation theory, outline a program of attack for the amenable cases, and summarize the present status of work on the staggered chiral forms and operators with good \sugts\ quantum numbers.}

\FullConference{XXIVth International Symposium on Lattice Field Theory\\
                July 23-28, 2006\\
                Tucson, Arizona, USA}

\begin{document}

\section{Introduction}
Successful calculations of the mass spectrum of the lightest octet and decuplet baryons would constitute an impressive benchmark for the staggered formulation of lattice QCD.  Ultimately, one might hope to include the effects of strong and electromagnetic isospin breaking and to develop staggered QCD into a powerful tool for determining the baryonic low-energy couplings (LECs) of chiral perturbation theory ($\chi$PT) and the quark model assignments of excited baryons.  Unfortunately, the staggered baryon spectrum is much more complicated than the physical spectrum; instead of three light quarks interacting in accord with \su{3}{F} symmetry, the valence sector contains twelve light quarks interacting in accord with \su{12}{f}.  In 2+1 flavor simulations, physical flavor breaking and taste violations lift the degeneracies of the \su{12}{f} multiplets and mix baryons that have the same conserved flavor-taste quantum numbers.  The resulting spectrum contains many nearly degenerate states that cannot be distinguished by their conserved quantum numbers~\cite{msip}; using 2+1 flavor staggered QCD to extract the masses of even the lightest baryons would be very difficult.  

However, in partially quenched simulations with three degenerate valence quarks, flavor \su{3}{F} ensures that certain of the troublesome splittings and mixings vanish.  As a direct consequence, high-precision calculations of the masses of the nucleon, the delta, and the omega may soon be possible.  Success requires improving the chiral extrapolations, developing a quantitative description of the taste violations, and avoiding the splittings and mixings in the spectrum.  The development of staggered $\chi$PT in the baryon sector addresses the first two concerns~\cite{dublin}.  The construction of operators transforming in irreps of \sugts\ addresses the third~\cite{msip}.  

\section{Staggered heavy baryon $\chi$PT and the staggered baryon spectrum}
Symanzik's effective continuum action for staggered QCD parameterizes the taste violations using irrelevant operators constructed from the elementary quark and gluon fields.  Mapping these operators into the corresponding hadronic operators yields the staggered chiral theory, which describes the taste violations in terms of the hadrons~\cite{leeshrp,aubin}.  Taste-changing processes involving a single external baryon can be described within the framework of heavy baryon $\chi$PT (HB$\chi$PT); the baryons are treated as non-relativistic, fixed-velocity sources, and observables are calculated in powers of the square roots of the up, down, and strange quark masses, the lattice spacing, and the average splitting of the lightest spin-\ohf\ and spin-\thf\ states.
\begin{equation}
\mathcal{O}(m_q^{1/2})=\mathcal{O}(a)=\mathcal{O}(\Delta)\;\;\;\;\;\;\;\;q=u,\,d,\,s \label{pwrcnt}
\end{equation}
The Lagrangian of staggered HB$\chi$PT (SHB$\chi$PT) needed to calculate the masses of the ground state spin-\ohf\ baryons to third order has been constructed, including the mass-dimension six operators from the Symanzik action needed for $\mc{O}(a^2)$ analytic contributions.  Generically, the baryon masses have the form
\begin{eqnarray}
M&=&const.+m_q+a^2+(m_q+a^2)^{3/2}\nonumber \\
&+&a^2(m_q+m_q^{1/2}a+a^2)^{1/2}+\Delta(m_q+a^2+\Delta^2)\ln(m_q+a^2)+\dots \label{genmass}
\end{eqnarray}

Because the hadrons of the staggered chiral theory transform in irreps of \su{12}{f}, the lightest spin-\ohf\ baryons transform in a \mb{572_M}, and the lightest spin-\thf\ baryons, in a \mb{364_S}.  To identify staggered baryons having the masses of physical baryons, one must choose a basis for these irreps in which the continuum limits of the masses of the staggered baryons are manifest.  Decomposing \su{12}{f} into irreps of \ft\ gives
\begin{eqnarray}
\mathbf{572_M}&\rightarrow&\mathbf{(10_S,\,20_M)\oplus(8_M,\,20_S)\oplus\;(8_M,\,20_M)\oplus(8_M,\,\bar 4_A)\oplus(1_A,\,20_M)}\label{572}\\
\mathbf{364_S}&\rightarrow&\mathbf{(10_S,\,20_S)\oplus(8_M,\,20_M)\oplus(1_A,\,\bar 4_A)}\label{364}
\end{eqnarray}
In the continuum limit, all members of a given \su{4}{T} multiplet are degenerate; the \su{3}{F} irreps suggest that, in the continuum limit, 44 baryons in the \mb{572_M} have the mass of a given baryon in the lightest physical octet, while 20 baryons in the \mb{364_S} have the mass of a given baryon in the decuplet.  In general, states in the \mb{(10_S,\,20_M)}, \mb{(1_A,\,20_M)}, and \mb{(1_A,\,\bar 4_A)}, as well as states in the \mb{(8_M,\,20_M)} of (\ref{364}), do not have physical masses.  However, in partially quenched simulations with degenerate valence quarks, all members of the \mb{572_M} are degenerate with the nucleon, while all members of the \mb{364_S} are degenerate with the partially quenched delta/omega---including baryons in product irreps with unphysical flavor irreps.  

\section{An example:  the flavor-symmetric nucleons in SHB$\chi$PT}
The corner-wall nucleon operator~\cite{golt,milc} interpolates to states in the \mb{(10_S,\,20_M)}.  For each member of the \mb{10_S}, the self-energy is a 20-dimensional matrix in baryon taste space.  To third order in the staggered chiral expansion, tree graphs contribute terms that are linear in the quark mass and the squared lattice spacing, loops with an intermediate spin-\ohf\ baryon, the first non-analytic terms, and loops with an intermediate spin-\thf\ baryon, the first chiral logarithms.  

In the rest frame of the heavy baryon formulation and to this order in the staggered chiral expansion, taste breaking occurs in two stages.  First the loops break taste to $\rt\subset \su{4}{T}$, and then the $\mc{O}(a^2)$ analytic terms break taste further to $\rrt\subset \rt$.  Under these remnant taste symmetries, the \mb{20_M} of \su{4}{T} decomposes:
\begin{equation}
\mb{20_M}\rightarrow\mb{12}\oplus2\mb{(4)}\rightarrow \mb{8}\oplus3\mb{(4)}\label{20dcmp}
\end{equation}
Taste violations in loops lift the continuum degeneracy of baryons in a \mb{12} and two \mb{4}s of \rt\ and introduce mixing between corresponding states in the two \mb{4}s; such states have the same conserved quantum numbers.  The form of the resulting mass matrix is dictated by charge conjugation and the \rt\ symmetry; in a suitable basis, the 8-dimensional submatrix corresponding to the two \mb{4}s (for any given member of the \mb{10_S}) has the form
\[\begin{pmatrix}
a & 0 & c & 0 & \cdots & & & \\
0 & a & 0 & -c & & & & \\
c & 0 & b & 0 & & & & \\
0 & -c & 0 & b & & & & \\
\vdots &  &  &  & a & 0 & c & 0 \\
 & & & & 0 & a & 0 & -c \\
 & & & & c & 0 & b & 0 \\
 & & & & 0 & -c & 0 & b
\end{pmatrix}\]
The elements in the suppressed off-diagonal blocks vanish.  The contributions parameterized here have been calculated to third order in the staggered chiral expansion, and this form of the mass matrix, explicitly verified.  The restoration of \su{4}{T} implies that $c$ must vanish in the continuum limit; a nontrivial but straightforward exercise with the loop contributions suffices to explicitly verify this consequence of taste symmetry.  

In the same way, taste violations in tree graphs of $\mc{O}(a^2)$ lift the \rt\ degeneracy of baryons in an \mb{8} and a \mb{4} of \rrt\ and introduce mixing between corresponding states in the three \mb{4}s of \rrt.  (Cf.~Eq.~(\ref{20dcmp}).)  The resulting submatrix containing nonzero off-diagonal elements is 12-dimensional and qualitatively similar to that shown above; although the $\mc{O}(a^2)$ elements are linear combinations of a finite number of LECs, explicit enumeration of the baryonic Lagrangian and calculation of these elements shows that no accidental degeneracies or other relations between them arise~\cite{msip}.  

\begin{figure}
\begin{center}
\includegraphics[width=0.7\textwidth]{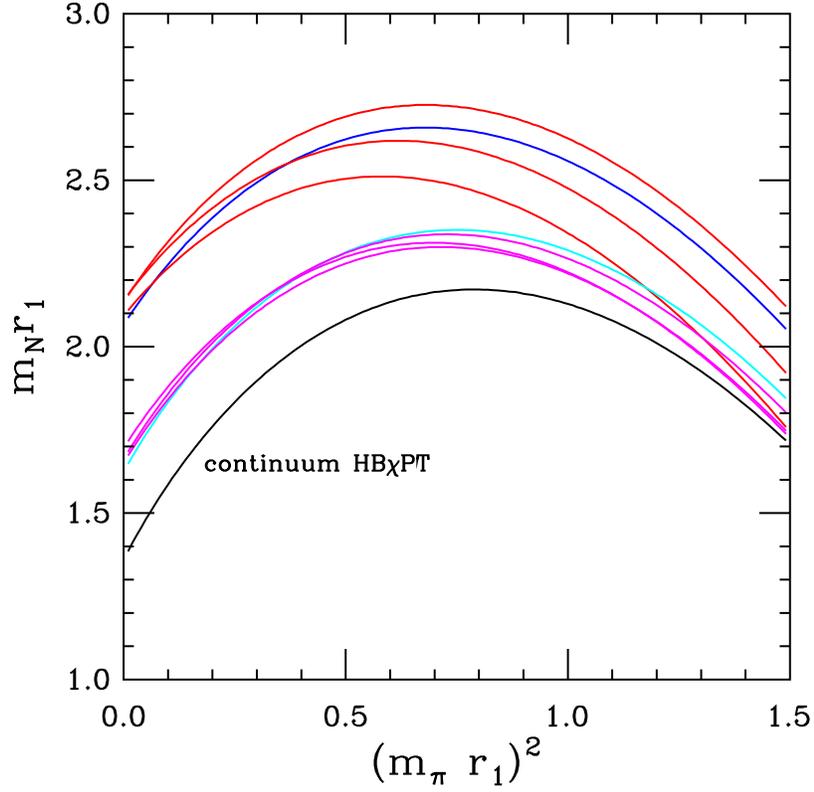}
\end{center}
\caption{Third order chiral forms for the masses of flavor-symmetric nucleons.  Red curves correspond to the three \mb{\color{red}4\color{black}}s of \rrt\ for a lattice spacing of about 0.12 fm.  Likewise, the blue curve corresponds to the \mb{\color{blue}8\color{black}} of \rrt.  The magenta and cyan curves are the same forms for a lattice spacing of about 0.09 fm; the black curve is the result in the continuum limit.  $r_1$ is about 0.317 fm.}
\label{chiforms}
\end{figure}

Fig.~\ref{chiforms} shows the nondegenerate diagonal elements of the mass matrix as functions of the squared continuum pion mass.  The curves are not fits but sketches constructed from the chiral forms by estimating the baryonic LECs and taking lattice parameters from the literature~\cite{dublin}.  At small up-down quark mass, the chiral forms rise linearly in accord with Eq.~(\ref{genmass}); however, at larger up-down masses, the forms decrease, while the lattice data increases monotonically with increasing up-down mass~\cite{dublin,milc}.  Although the fact that the forms eventually decrease is not a surprise---for large quark masses, they behave like the negative of the meson mass cubed---this decrease strongly suggests that third order staggered $\chi$PT cannot adequately describe the data.  At sufficiently large up-down masses, the staggered chiral power counting of Eq.~(\ref{pwrcnt}) breaks down, and continuum chiral corrections become much more important than the taste violations.  If the data at the larger up-down masses can be adequately described by continuum chiral corrections, then one could improve control over the continuum extrapolations even while neglecting fourth-order taste violations.  

To arrive at Fig.~\ref{chiforms}, one must estimate the magnitude of the $\mc{O}(a^2)$ corrections.  Their natural size is set by the fact that they absorb the scale dependence of the spin-\thf\ loops; one finds that the splitting between the \rrt\ multiplets ranges from about 10 to 40 MeV, an estimate that depends on the lattice spacing and quark masses.  The first excited states are therefore probably quite close to the ground state; although presenting no problems of principle, this situation makes excited state contamination a significant source of systematic error and a difficult practical problem.  One is thus led to consider the operators that create each of these states.  

\section{Operators for the single-flavor nucleons}
Consider interpolating fields for the single-flavor members of the \mb{(10_S,\,20_M)}; such operators transform within three irreps of the geometrical time slice group, \mr{GTS}:  the \mb{8}, the \mb{8^{\prime}}, and the \mb{16}, where the two 8-dimensional irreps are inequivalent~\cite{golt}.  Decomposing the $\su{2}{S}\times \su{4}{T}$ irrep of the single-flavor nucleons into irreps of $\mr{GTS}\subset \su{2}{S}\times \su{4}{T}$ gives \[\mb{({\textstyle\frac{1}{2}}, \,20_M)}\rightarrow\mathbf{\color{blue}16\color{black}}\oplus 3\mathbf{(\color{red}8\color{black})},\]
which implies that operators transforming in \mb{8}s of \mr{GTS} create three single-flavor nucleons, and operators in the \mb{16}, one.  

To associate the operators with the appropriate chiral forms, consider how taste violations break the continuum $\su{2}{S}\times \su{4}{T}$ to $\mr{GTS}\subset \su{2}{S}\times [\rrt]\subset \su{2}{S}\times \su{4}{T}$:  
\[\mathbf{(\textstyle{\frac{1}{2}},\,20_M)}\rightarrow \mathbf{(\textstyle{\frac{1}{2}},\,\color{blue}8\color{black})}\oplus 3\mathbf{(\textstyle{\frac{1}{2}},\,\color{red}4\color{black})}\rightarrow \mb{\color{blue}16\color{black}}\oplus 3\mb{(\color{red}8\color{black})}\]
The decomposition of the \mb{20_M} into irreps of \rrt\ is the same as the second decomposition shown in Eq.~(\ref{20dcmp}); the decomposition into irreps of \mr{GTS} reflects the fact that the $\su{2}{S}\times [\rrt]$ irreps are already irreducible under \mr{GTS}.  Operators transforming in \mb{8}s of \mr{GTS} create corresponding nucleons in all three \mb{4}s of Fig.~\ref{chiforms}, while operators transforming in \mb{16}s create nucleons in the lone \mb{8}.  

To summarize, operators transforming in the \mb{8} of \mr{GTS} create three closely spaced states that become degenerate in the continuum limit with the nucleon.  Corresponding nucleons in the two excited multiplets not only have nearly the same mass as nucleons in the ground state multiplet, but also have the same conserved quantum numbers and therefore mix with the ground state nucleons and one another as discussed in the previous section.  In contrast, operators transforming in the \mb{16} of \mr{GTS} create nucleons in only one multiplet, so such operators could be used to circumvent the difficulties that would be involved in accounting for the excited states in the spectrum created by operators transforming in the \mb{8}.  

Although this conclusion is heartening, it was reached without considering the spin-\thf\ states created by operators in the \mb{16}.  An argument analogous to that given in this section shows that, for extracting the masses of the nucleon and the delta/omega, certain operators with good \su{3}{F} quantum numbers are preferable to operators in the \mb{16}.  

\section{Operators with good \su{3}{F} quantum numbers}
Consider the spectra created by operators transforming in irreps of \sugts\ when such operators are used in partially quenched simulations with degenerate valence quarks.  In the continuum limit, the lightest spin-\ohf\ states are all degenerate with the nucleon, while the lightest spin-\thf\ states, with the delta/omega.  Decomposing the \spinks\ irreps into irreps of $\sugts\subset\spinks$ gives
\begin{eqnarray*}
\mb{(\ohf,\,572_M)}&\rightarrow & 3\mb{(10_S,\,\color{red}8\color{black})\oplus(10_S,\,\color{blue}16\color{black})}\oplus5\mb{(8_M,\,8)}\oplus3\mathbf{(8_M,\,16)}\oplus3\mathbf{(1_A,\,8)\oplus(1_A,\,16)} \\
\mb{(\thf,\,364_S)}&\rightarrow & 2\mb{(10_S,\,\color{red}8\color{black})}\oplus2\mb{(10_S,\,8^{\prime})}\oplus 3\mb{(10_S,\,\color{blue}16\color{black})}\oplus \mb{(8_M,\,8)}\oplus\mathbf{(8_M,\,8^{\prime})}\oplus4\mathbf{(8_M,\,16)}\oplus\mathbf{(1_A,\,16)}
\end{eqnarray*}
%The implied spectra of these operators are shown in Fig.~\ref{spectra}.
In accord with the discussion of the single-flavor case, the \mb{(10_S,\,8)} occurs three times in the decomposition of the \mb{(\ohf,\,572_M)}, and the \mb{(10_S,\,16)}, once.  However, these irreps both occur more than once in the decomposition of the \mb{(\thf,\,364_S)}; the \mb{(10_S,\,16)} appears three times, so operators transforming in this irrep (including the single-flavor operators) create three nearly degenerate spin-\thf\ states with the same conserved quantum numbers.  If one wants to account for delta/omega contamination, then the difficulties associated with the splittings and mixings reassert themselves.  
%\begin{figure}
%\begin{center}
%\includegraphics[width=0.8\textwidth]{spectra_eps2.ps}
%\end{center}
%\caption{Schematic of the \sugts\ operator spectra for partially quenched simulations with degenerate valence quarks and $m_u=m_d$ in the sea.  States above the dotted line become degenerate, in the continuum limit, with the delta/omega; states below, with the nucleon.}
%\label{spectra}
%\end{figure}

However, two \sugts\ irreps appear at most once in each decomposition.  The \mb{(1_A,\,16)} occurs once in each, while the \mb{(8_M,\,8^{\prime})} occurs only in the decomposition of the \mb{(\thf,\,364_S)}.  Therefore, operators transforming in the \mb{(1_A,\,16)} create a single nucleon and a single delta/omega, while operators transforming in the \mb{(8_M,\,8^{\prime})} create only a lone delta/omega.  Although the nucleon and the delta/omega created by operators in the \mb{(1_A,\,16)} do mix, a mass difference of about 200 MeV indicates that numerical analysis should be comparatively feasible.  

In summary, in partially quenched simulations with degenerate valence quarks, two linearly independent operators transforming in \mb{(1_A,\,16)}s of \sugts\ could be used to extract the masses of the nucleon, the delta, and the omega.  The masses of the delta and omega could also be extracted using an operator transforming in the \mb{(8_M,\,8^{\prime})}.  Operators transforming in these irreps and the other \sugts\ irreps appearing above have been constructed, and chiral forms for the nucleons and delta/omegas created by operators in the \mb{(1_A,\,16)} and \mb{(8_M,\,8^{\prime})} have been calculated to third order in SHB$\chi$PT~\cite{msip}.  For example, one operator transforming in a \mb{(1_A,\,16)} is 
\begin{eqnarray*}
\sum _{\mathbf{x},\,x_k\;\mathrm{even}}\textstyle{\frac{1}{6}}\epsilon^{ijk} \textstyle{\frac{1}{6}}\epsilon_{abc}&\Biggl\{\sum _{\varepsilon_2\varepsilon_3} \chi_i^a(\mathbf{x}+\mathbf{a}_1)[U(\mathbf{x}+\mathbf{a}_1,\,\mathbf{x}+\mathbf{a}_1+\varepsilon_2\mathbf{a}_2)\chi_j(\mathbf{x+a}_1+\varepsilon_2\mathbf{a}_2)]^b\times&\\
&[U(\mathbf{x+a}_1,\,\mathbf{x+a}_1+\varepsilon_3\mathbf{a}_3)\chi_k(\mathbf{x+a}_1+\varepsilon_3\mathbf{a}_3)]^c&\\
&-\sum _{\varepsilon_1\varepsilon_3}\chi_i^a(\mathbf{x+a}_2)[U(\mathbf{x+a}_2,\,\mathbf{x}+\varepsilon_1\mathbf{a}_1+\mathbf{a}_2)\chi_j(\mathbf{x}+\varepsilon_1\mathbf{a}_1+\mathbf{a}_2)]^b\times&\\
&[U(\mathbf{x+a}_2,\,\mathbf{x+a}_2+\varepsilon_3\mathbf{a}_3)\chi_k(\mathbf{x+a}_2+\varepsilon_3\mathbf{a}_3)]^c\Biggr\},&
\end{eqnarray*}
where $ijk$ are \su{3}{F} indices, $abc$ are color indices, and the remaining notation is defined in Ref.~\cite{golt}.  This particular operator transforms in an $E_{-1}^{-}$ irrep of the octahedral subgroup of \mr{GTS}.  Gauge links have been added to make it gauge invariant~\cite{golt}.  

\section{Summary}
The masses of the flavor-symmetric nucleons have been calculated to third order in staggered, partially quenched HB$\chi$PT.  The results for the diagonal and off-diagonal elements in the mass matrix have been checked against the pattern of degeneracies implied by the remnant taste symmetries of the loops and $\mc{O}(a^2)$ contributions.  In the continuum limit, all off-diagonal elements of the mass matrix explicitly vanish because taste \su{4}{T} is restored, while the diagonal elements reduce to the results obtained from continuum partially quenched HB$\chi$PT.

The masses of the flavor-antisymmetric nucleon, the flavor-antisymmetric delta/omega, and the flavor-mixed delta/omega have been calculated to third order in staggered, partially quenched HB$\chi$PT.  To this order no mixing occurs between the spin-\ohf\ and spin-\thf\ flavor-antisymmetric states, which are created by interpolating fields transforming in the irrep \mb{(1_A,\,16)} of \sugts.  The flavor-mixed delta/omega is created by interpolating fields transforming in the irrep \mb{(8_M,\,8^{\prime})} of \sugts.  Operators transforming in these irreps and in the other \sugts\ irreps have been constructed~\cite{msip}.  Current work includes coding the operators transforming in the \mb{(1_A,\,16)} and \mb{(8_M,\,8^{\prime})}.  

Using the entire set of operators, one could, in principle, extract the masses of all the lightest spin-\ohf\ and spin-\thf\ baryons of staggered QCD.  However, the practical difficulties associated with the splittings and mixings, which proliferate when one breaks \su{3}{F} in the valence sector, would make such a program very difficult.


\begin{thebibliography}{99}
  \bibitem{msip} Jon~A.~Bailey, manuscripts in preparation.
  \bibitem{dublin} Jon~A.~Bailey and C.~Bernard, in proceedings of the \emph{XXIIIrd Int. Symp. on Lattice Field Theory}, PoS(LAT2005)047 [hep-lat/0510006].
  \bibitem{leeshrp} W.~J.~Lee and S.~R.~Sharpe, Phys. Rev. D {\bf 60}, 114503 (1999) [{{hep-lat/9905023}}].
  \bibitem{aubin} C.~Aubin and C.~Bernard, Phys. Rev. D {\bf 68}, 034014 (2003) [{{hep-lat/0304014}}] and 074011 (2003) [{{hep-lat/0306026}}]; C.~Bernard, Phys. Rev. D {\bf 65}, 054031 (2002).
  \bibitem{golt} M.~F.~L.~Golterman and J.~Smit, Nucl. Phys. {\bf B 255}, 328-340 (1985).
  \bibitem{milc} C.~Aubin \emph{et al.}, Phys. Rev. D {\bf 70}, 094505 (2004) [{{hep-lat/0402030}}].
%  \bibitem{labrenz} J.~N.~Labrenz and S.~R.~Sharpe, Phys. Rev. D \bf{54}, 4595 (1996) [{{hep-lat/9605034}}].
%  \bibitem{andre} A.~Walker-Loud, Nucl. Phys. \bf{A 747}, 476-507 (2005) [{{hep-lat/0405007}}].
\end{thebibliography}
\end{document}